\documentclass[floats,floatfix,showpacs,amssymb,prd,twocolumn,superscriptaddress,nofootinbib,nolongbibliography,reprint]{revtex4-1}

\usepackage{amssymb,amsmath,verbatim,mathtools,needspace,enumitem,etoolbox,graphicx,physics,microtype,afterpage,xspace,tabularx,lmodern,multirow}
\usepackage{gensymb}
\usepackage[normalem]{ulem}
\usepackage[dvipsnames, usenames]{xcolor}
\definecolor{linkcolor}{rgb}{0.0,0.3,0.5}
\usepackage[unicode, colorlinks=true, linkcolor=linkcolor, citecolor=linkcolor, filecolor=linkcolor, urlcolor=linkcolor, linktocpage, breaklinks]{hyperref}
\usepackage[all]{hypcap}
\usepackage{subfigure}
\usepackage[T1]{fontenc}
\usepackage[utf8]{inputenc}
\usepackage{diagbox} %
\usepackage{tikz}
\newcommand\diag[4]{%
  \multicolumn{1}{p{#2}|}{\hskip-\tabcolsep
  $\vcenter{\begin{tikzpicture}[baseline=0,anchor=south west,inner sep=#1]
  \path[use as bounding box] (0,1) rectangle (#2+2\tabcolsep,\baselineskip);
  \node[minimum width={#2+2\tabcolsep},minimum height=\baselineskip+17] (box) {};
  \draw (box.north west) -- (box.south east);
  \node[anchor=south west] at (box.south west) {#3};
  \node[anchor=north east] at (box.north east) {#4};
 \end{tikzpicture}}$\hskip-\tabcolsep}}
\usepackage[usenames,dvipsnames]{xcolor}
\hypersetup{colorlinks=true,citecolor=romared,linkcolor=romared,urlcolor=romared}

\definecolor{romared}{RGB}{142,0,28}

\newcommand{\be}{\begin{equation}}
\newcommand{\ee}{\end{equation}}

\def\be{\begin{equation}}
\def\ee{\end{equation}}
\newcommand{\beq}{\begin{eqnarray}}
\newcommand{\eeq}{\end{eqnarray}}

\usepackage{aas_macros}
\usepackage{makecell}
\usepackage{soul}
\usepackage{amssymb}

\usepackage{lipsum}

\newcolumntype{Y}{>{\centering\arraybackslash}X}

\begin{document}
\title{Black holes surrounded by generic matter distributions: \\ polar perturbations and energy flux}

\begin{abstract}
  We develop a numerical approach to compute polar parity perturbations within fully relativistic models of black hole systems embedded in generic, spherically symmetric, anisotropic fluids. We apply this framework to study gravitational wave generation and propagation from extreme mass-ratio inspirals in the presence of several astrophysically relevant dark matter models, namely the Hernquist, Navarro-Frenk-White, and Einasto profiles. We also study dark matter spike profiles obtained from a fully relativistic calculation of the adiabatic growth of a BH within the Hernquist profile, and provide a closed-form analytic fit of these profiles. Our analysis completes prior numerical work in the axial sector, yielding a fully numerical pipeline to study black hole environmental effects. We study the dependence of the fluxes on the DM halo mass and compactness. We find that, unlike the axial case, polar fluxes are not adequately described by simple gravitational-redshift effects, thus offering an exciting avenue for the study of black hole environments with gravitational waves.
\end{abstract}

\newcommand{\jhu}{William H.\ Miller III Department of Physics and Astronomy, Johns Hopkins University, \\ 3400 N. Charles Street, Baltimore, Maryland, 21218, USA}
\newcommand{\GSSI}{Gran Sasso Science Institute (GSSI), I-67100 L’Aquila, Italy}
\newcommand{\GranSasso}{INFN, Laboratori Nazionali del Gran Sasso, I-67100 Assergi, Italy}

\author{Nicholas Speeney}
\email{nspeene1@jhu.edu}
\address{\jhu}

\author{Emanuele Berti}
\email{berti@jhu.edu}
\address{\jhu}

\author{Vitor Cardoso} 
\email{vitor.cardoso@tecnico.ulisboa.pt}
\address{Niels Bohr International Academy, Niels Bohr Institute, Blegdamsvej 17, 2100 Copenhagen, Denmark}
\address{CENTRA, Departamento de F\'{\i}sica, Instituto Superior T\'ecnico -- IST, Universidade de Lisboa -- UL, Avenida Rovisco Pais 1, 1049-001 Lisboa, Portugal}
\address{Yukawa Institute for Theoretical Physics, Kyoto University, Kyoto}

\author{Andrea Maselli}
\email{andrea.maselli@gssi.it}
\address{\GSSI}
\address{\GranSasso}

\date{\today}
\maketitle

\section{Introduction}
\label{sec:intro}

Astrophysical black hole (BH) binaries are, in general, nonvacuum systems. The environment surrounding a BH binary is expected to play an important role in the generation and propagation of gravitational waves (GWs) by modifying the orbital trajectory of the binary and the GW phase~\cite{Barausse:2014tra,Yunes:2011ws,Cardoso:2019rou}. These modifications are particularly interesting because the properties of the environment are encoded in the gravitational radiation. The relation between the GW signal and the properties of the environment has been studied in many cases of interest, including accretion disks~\cite{Yunes:2011ws,Speri:2022upm}, dark matter (DM) clouds and overdensities~\cite{Eda:2013gg,Eda:2014kra,Boskovic:2018rub,Annulli:2020lyc,Coogan:2021uqv,Traykova:2021dua,Cole:2022fir,Speeney:2022ryg}, bosonic clouds~\cite{Macedo:2013qea,Baumann:2021fkf,Barsanti:2022vvl,Barsanti:2022ana,Vicente:2022ivh,Duque:2023cac}, etcetera.

All of these prior studies are based on simplifying assumptions. Very often, the effect of the environment is modeled by including Newtonian corrections to the GW quadrupole formula (see e.g.~\cite{Eda:2013gg,Eda:2014kra}). In the few cases in which the environment is treated relativistically~\cite{Speeney:2022ryg,Vicente:2022ivh}, the orbital trajectory and GW emission are modeled within the post-Newtonian (PN) approximation, a perturbative expansion in the ratio $v/c$ of the binary's orbital velocity to the speed of light.

Extreme mass-ratio inspirals (EMRIs) of stellar-mass objects with mass $m_p$ around massive BHs, with mass ratios $q=m_p/M_{\rm BH} \sim \mathcal{O}(10^{-5}-10^{-8})$, can spend $\sim 10^5$ orbital cycles in the low-frequency regime accessible to space-based detectors such as the Laser Interferometer Space Antenna (LISA)~\cite{LISA:2017pwj}, and they are expected to lead to the best constraints on environmental effects~\cite{Cardoso:2019rou}.

The approximations commonly used in the treatment of environmental effects are not accurate enough for EMRIs. These binaries are not well modeled within the PN framework, which is more amenable to comparable mass systems~\cite{Yunes:2008tw,Zhang:2011vha,Sago:2016xsp}. In addition, these studies do not consistently include the backreaction effect of the environment on the background solution (which is also expected to modify the orbital dynamics and the GW signal) in a fully relativistic manner.
It is therefore imperative to correctly model the environment and the orbital dynamics as precisely as possible to fully capture secular effects which may accumulate throughout the orbit. 

Recent work has established a fully relativistic approach to address the need for accurate modeling of EMRIs embedded in astrophysical environments. This program was initiated in Ref.~\cite{Cardoso:2021wlq}, which employed an extension of the Einstein cluster model to find relativistic solutions for the background spacetime and to compute observable quantities. Later work established a complete framework to study the coupled gravitational and fluid perturbations of either parity (axial or polar) for generic, spherically symmetric density distributions affecting EMRI inspirals~\cite{Cardoso:2022whc}. While the framework developed in these works is generic, the analysis focused on the Hernquist distribution~\cite{Hernquist:1990be}, which yields closed-form analytic results for the background metric. Further progress was made in Ref.~\cite{Figueiredo:2023gas}, which developed a fully numerical pipeline to treat generic density distributions in the axial sector.

In this paper we complete this research program by extending the analysis of Ref.~\cite{Figueiredo:2023gas} to include polar perturbations. Polar perturbations in the presence of generic density profiles can be reduced to 5 frequency-domain ordinary differential equations that couple the variables describing perturbations of the fluid with the variables describing gravitational perturbations. This is different from the axial case, in which the gravitational and fluid variables are decoupled, and therefore they can be analyzed separately. The framework we develop here is valid for any spherically symmetric distribution surrounding the central BH, but we focus on a set of specific DM models (including the Hernquist, Navarro-Frenk-White, and Einasto profiles) which are commonly used to describe the DM environment affecting the evolution of the EMRI. We compute the polar GW energy flux multipoles for various DM models and configurations of the halo, and then we compare these calculations with previously computed axial multipoles. Throughout the paper we use geometrical units ($G=c=1$).

\section{Background and Density Profiles}
In this section we review the formalism used to construct relativistic BH solutions embedded within a DM halo. We consider a spherically symmetric, static spacetime, described by the line element
\begin{align}
ds^2=&g^{(0)}_{\mu \nu}dx^\mu dx^\nu\nonumber\\
=&-a(r)dt^2+\frac{dr^2}{1-2m(r)/r}+r^2d\Omega^2\ ,
\end{align}
which solves the nonvacuum Einstein equations
\begin{equation}
\label{eq:background_Einstein}
G_{\mu \nu}=8\pi T^{\text{env}}_{\mu \nu}\ .
\end{equation}

Following the Einstein cluster prescription~\cite{Einstein:1939ms,Geralico:2012jt}, the properties of the environment are described by an anisotropic stress-energy tensor $(T^{\text{env}})^\mu{_\nu}=\text{diag}(-\rho,0,P_t,P_t)$, where $\rho(r)$ and $P_t(r)$ are the density and the tangential pressure of the matter distribution.
The continuity equation determines the mass profile through the relation $m'(r)=4\pi r^2 \rho(r)$. The metric functions $a(r)$ and $P_t(r)$ are found from the $rr$ component of Eqs.~\eqref{eq:background_Einstein} and from the Bianchi identities, respectively:
\begin{equation}
\label{eq:background_eqs}
\frac{a'(r)}{a(r)}=\frac{2m(r)/r}{r-2m(r)}\quad\ ,\
\quad P_t(r)=\frac{m(r)/2}{r-2m(r)}\rho(r)\ .
\end{equation}
The solutions of Eqs.~\eqref{eq:background_eqs} and the function $m(r)$, for a given profile $\rho(r)$, fully specify the background spacetime. These equations can be integrated numerically~\cite{Figueiredo:2023gas}.

The form of $m(r)$ and $a(r)$ also determines the properties of particle geodesics in the spacetime. The energy and the angular momentum of a particle on a circular orbit with radius $r_p$ are given by
\begin{equation}
\label{eq:enr_ang}
E_p=\left[ \frac{r-2m(r)}{r-3m(r)}a(r) \right]^{1/2}_{r=r_p}\ , ~~~ L_p=\left[ \frac{m(r)}{r-3m(r)} \right]^{1/2}_{r=r_p}\ ,
\end{equation}
while the orbital frequency is $\Omega_p= a(r_p) L_p/r_p^2 E_p$. Hereafter, without loss of generality, we assume that the orbits are equatorial and set $\theta=\pi/2$.

With the background spacetime solution in hand, we can now examine the orbital dynamics of EMRIs and assess the relevance of the environment on the resulting GW emission. At the leading dissipative order, the secondary can be treated as a particle of mass $m_p$ inducing small perturbations of the background metric and of the stress-energy tensor, such that
\begin{align}
\label{eq:Perturbed_metric}
 g_{\mu \nu}=g_{\mu \nu}^{(0)}+g_{\mu \nu}^{(1)}\quad \ ,
\quad T_{\mu \nu}^{\text{env}}=T_{\mu \nu}^{(0)\text{env}}+T_{\mu \nu}^{(1)\text{env}} \ ,
\end{align}
and
\begin{equation}
\label{eq:Perturbed_Einstein}
G_{\mu \nu}^{(1)}=8\pi T_{\mu \nu}^{(1)\text{env}}+8\pi T_{\mu \nu}^p\,,
\end{equation}
where $T_{\mu \nu}^p$ is the stress-energy tensor of the particle. The metric and matter perturbations can be decomposed in terms of axial and polar tensor spherical harmonics, yielding decoupled radial equations due to the symmetries of the background spacetime~\cite{Regge:1957td,Zerilli:1970wzz,Lindblom:1983ps}.

The axial and polar equations have been derived in Ref.~\cite{Cardoso:2022whc}, and for brevity we do not reproduce them here. The axial perturbations have been extensively investigated in Ref.~\cite{Figueiredo:2023gas} for both analytical and numerical backgrounds, assuming different DM profiles. The polar sector has been studied in Ref.~\cite{Cardoso:2022whc} for the Hernquist model, which has the advantage that the background spacetime can be written down in a closed analytic form.

\subsection{Dark matter density profiles}
In this work we focus on polar perturbations, with the goal of extending previous calculations through a fully numerical pipeline capable of describing generic DM profiles.

\begin{figure*}[t]
  \centering
  \includegraphics[width=\linewidth]{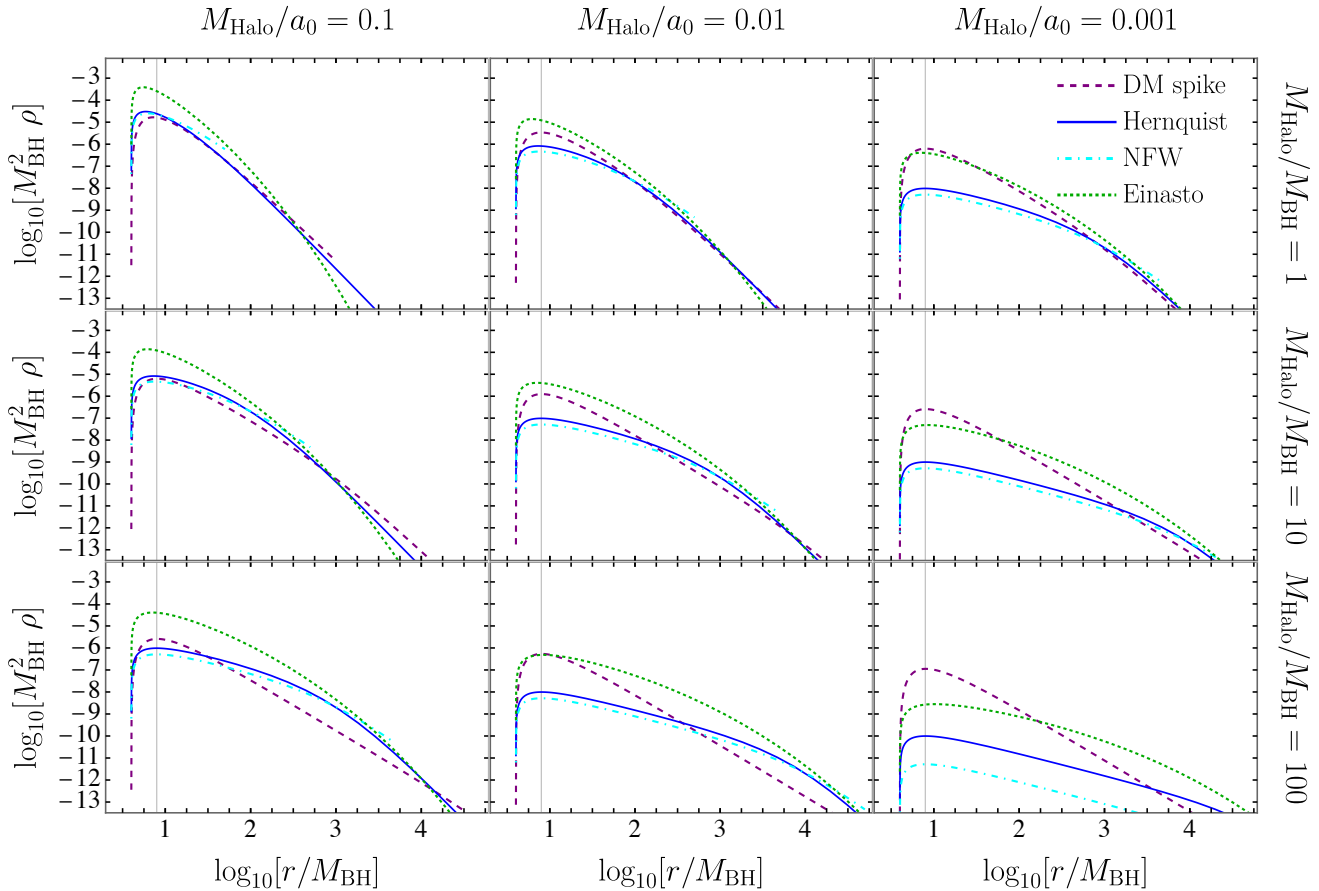}
  \caption{Mass profiles for some representative DM configurations considered in this work. The NFW model is truncated at the cutoff radius $r_c=5a_0$. The Hernquist, NFW and Einasto profiles are obtained by the ``cutoff factor'' recipe $\rho(r)\rightarrow (1-4M_{\text{BH}}/r)\rho(r)$. The ``DM spike'' profile is obtained with the fully relativistic calculation of the adiabatic growth of a BH within a Hernquist profile from Refs.~\cite{Sadeghian:2013laa,Speeney:2022ryg}, as described in the main text. Going from the top to the bottom panels, the DM halo mass grows relative to the central BH mass ($M_{\text{Halo}}/M_{\text{BH}}=1,\,10,\,100$), while the compactness of the halo decreases going from the left to the right panels ($M_{\text{Halo}}/a_0=0.1,\,0.01,\,0.001$). The vertical lines correspond to a representative orbital radius $r_p=7.9456M_{\text{BH}}$ that will be used in Table~\ref{tab:fluxes} below to compare GW energy fluxes.}
\label{fig:Density_profiles_grid}
\end{figure*}

We consider a selection of DM distributions proposed to model cold DM profiles within galaxies.  Among them, the Hernquist \cite{Hernquist:1990be} and Navarro-Frenk-White (NFW)~\cite{Navarro:1996gj} distributions, which have been widely studied and compared to both observations and N-body simulations, can be parametrically described as
\begin{equation}
\label{eq:Hernquist_NFW_Density}
\rho(r)=\rho_0(r/a_0)^{-\gamma} [1+(r/a_0)^\alpha]^{(\gamma-\beta)/\alpha}\ ,
\end{equation}
where $a_0$ is the scale radius of the halo and $\rho_0$ is a scale factor proportional to the density through the relation $\rho_0=2^{(\beta-\gamma)/\alpha}\,\rho(a_0)$. The exponents $\gamma$ and $\beta$ control the slope of the distribution at small and large radii, respectively.  The values of $a_0$ and $\alpha$ determine the location and ``sharpness'' of the change in slope. By fixing $(\alpha,\beta,\gamma)=(1,4,1)$ and $(\alpha,\beta, \gamma)=(1,3,1)$ we recover the Hernquist profile and the NFW profile, respectively.  Note that the total mass of the NFW distribution is logarithmically divergent. Therefore, we define a cutoff radius $r_c$ such that $M_{\text{Halo}}(r>r_c)=0$ (as in, for example, Ref.~\cite{Figueiredo:2023gas}).

We also consider a third, observationally driven model: the Einasto profile, commonly used to study the galaxy luminosity distribution~\cite{Einasto:1965czb,1969Afz.....5..137E} and such that
\begin{equation}
\label{eq:Einasto_density}
\rho(r)=\rho_e \exp\left\{-d_n[(r/r_e)^{1/n}-1]\right\}\ ,
\end{equation}
with $n=6$ and $d_n=53/3$ \cite{Graham:2005xx,Prada:2005mx}.  Here $\rho_e$ is the density at radius $r_e$, which defines a volume containing half of the halo mass. Throughout this work, we set $r_e=a_0$.

The DM density profile arising from the (adiabatic) accretion growth of a nonrotating BH sitting at the core is expected to develop an overdense cusp followed by a sharp cutoff close to the horizon~\cite{Gondolo:1999ef}.  A general relativistic analysis predicts that the density falls off at $r=4M_{\text{BH}}$~\cite{Sadeghian:2013laa}. In our numerical profiles, we mimic this behavior by considering two approaches. The first approach consists of following a strategy similar to that adopted in Ref.~\cite{Figueiredo:2023gas}, i.e., we multiply the density of the DM models discussed above by a ``cutoff factor'' such that $\rho(r)\rightarrow (1-4M_{\text{BH}}/r)\rho(r)$. The prefactor of $(1-4M_{\text{BH}}/r)$, rather than the prefactor of $(1-2M_{\text{BH}}/r)$ used in in Ref.~\cite{Figueiredo:2023gas}, is chosen to avoid potential problems with the tangential pressure which arise in the region $2M_{\text{BH}}<r<4M_{\text{BH}}$ \cite{Datta:2023zmd}: with the present choice, the ratio $P_t(r)/\rho(r)$ remains finite and smaller than unity for all $r$. In the second approach we consider the fully relativistic ``DM spike'' models investigated by some of us in previous work~\cite{Speeney:2022ryg}. In that work, following Ref.~\cite{Sadeghian:2013laa}, we derived semianalytic fits valid for $r\lesssim a_0$ for spiked profiles based on the Hernquist and NFW distributions, and used these fits to assess the impact of the overdensities on the GW emission of binary sources.

From now on we will focus on the Hernquist subclass of the DM spikes, i.e., we consider the adiabatic growth of a spike starting from Eq.~\eqref{eq:Hernquist_NFW_Density} with $(\alpha,\beta,\gamma)=(1,4,1)$. This is because the Hernquist and NFW DM spikes are very similar at small values of $r$ (see Fig.~\ref{fig:Density_profiles_grid}).

We further improve the fits of the Hernquist DM spike profile from Ref.~\cite{Speeney:2022ryg} and extend their domain of validity to large radii (i.e., to the domain $r\gg a_0$). The result is of the form
\begin{equation}
\label{eq:DM_spike_fit}
\rho(r)=\frac{M_{\text{Halo}}}{M_{\text{BH}}^2}\left(\int_{4M_{\text{BH}}}^{r_c}4\pi r^2 \bar{\rho}(r) dr \right)^{-1} \bar{\rho}(r)
\end{equation}
with
\begin{equation}
\bar{\rho}(r)=\left(1-\frac{4M_{\text{BH}}}{r} \right)^\alpha \left( \frac{r M_{\text{BH}}}{M_{\text{Halo}}a_0}\right)^\beta \left(1+\frac{r M_{\text{BH}}}{M_{\text{Halo}}a_0} \right)^\gamma\ ,
\end{equation}
where $\alpha=2.366$, $\beta=-2.320$, and $\gamma=-1.370$.  The functional dependence of $\bar{\rho}(r)$ on $M_{\text{Halo}}$ and $a_0$ was determined empirically by studying the behavior of an ensemble of DM profiles computed numerically for selected values of $M_{\text{Halo}}$ and $a_0$~\cite{Sadeghian:2013laa}. The values of the exponents $(\alpha,~\beta,~\gamma)$ were then found by fitting with \texttt{Mathematica} a numerical spike with $M_{\text{Halo}}=10^4M_{\text{BH}}$ and $M_{\text{Halo}}/a_0=0.001$.

In the calculation of the GW fluxes (see Sec.~\ref{sec:Numerical_Procedure} below) we enforce a cutoff radius of $r_{c}=100M_{\text{Halo}}a_0/M_{\text{BH}}$ such that $M_{\text{Halo}}(r>r_c)=0$ for the DM spikes.
Equation~\eqref{eq:DM_spike_fit} reproduces the numerical results with $2\%$ accuracy in the range $r_{\text{ISCO}}<r<r_c$ for the values of $(M_{\text{Halo}}/M_{\text{BH}},M_{\text{Halo}}/a_0)$
which we investigated in this work.
The accuracy of the fit generally improves as both $M_{\text{Halo}}$ and $a_0$ increase.

In Fig.~\ref{fig:Density_profiles_grid} we plot some of the DM density distributions described so far for representative choices of the halo parameters.
The Hernquist and NFW profiles are always rather similar, except at the highest values of the halo mass and smallest values of the compactness (bottom right panel), where the density of the Hernquist model is larger than the NFW density by about one order of magnitude.
When the halo is very compact (left panels) the Einasto profiles have the highest DM density close to the BH, orders of magnitude larger than the DM spike profiles. The trend reverses as we move to the less compact profiles on the right: for the largest halo masses and for the smallest values of the compactness (bottom right panel) the DM spike contains by far the largest amount of DM at the orbital radii of interest for EMRIs.

Note that in all the cases we consider, the DM halo density and compactness are quite high compared to astrophysical galactic halos. Observations 
of SgrA$^*$ at the Galactic centre 
from the GRAVITY collaboration show that 
the extended mass in the region $r\lesssim10^6M_{\text{BH}}$ is very small, with the total mass function dominated by the central $4.3\times 10^6 M_\odot$ BH~\cite{GRAVITY:2021xju}. The distributions considered in this work have masses $\sim(1-10)M_{\text{BH}}$ within radii $r<10^4M_{\text{BH}}$, and they were mainly chosen to provide a proof of concept that our computational pipeline can be used to study relativistic environmental effects. These distribution may still be representative of DM halos in galaxies smaller than the Milky Way, or of subhalos~\cite{Springel:2008cc}.

\section{Numerical Procedure}
\label{sec:Numerical_Procedure}
We outline here the numerical procedure followed to compute the polar gravitational perturbations.

We first make a specific choice of $\rho(r)$, and then we solve for the background metric component $a(r)$ and for the mass function $m(r)$. In lieu of analytic or approximate solutions as found in \cite{Shen:2023erj,Konoplya:2022hbl}, we choose to numerically solve for the metric components by numerically integrating $m'(r)=4\pi r^2 \rho(r)$ from $r=2M_{\text{BH}}$ to the cutoff  radius $r_{\infty}=10^{11}M_{\text{BH}}$, which corresponds to our numerical approximation of spatial infinity. We then plug the numerical solution for $m(r)$ into Eq.~\eqref{eq:background_eqs} to solve for $a(r)$ by integrating ``backwards'' from $r_{\infty}$, with the following boundary condition:
\begin{align}
\label{eq:a(r)_inf_BC}
m(r_{\infty})=&M_{\text{BH}}+M_{\text{Halo}}\ ,\nonumber\\
a(r_{\infty})=&1-\frac{2m(r_{\infty})}{r_{\infty}}\ .
\end{align}
Here, $M_{\text{Halo}}$ corresponds to the total mass of the environment surrounding the BH.

The numerical solutions for $a(r)$ and $m(r)$ are used to calculate the tangential pressure $P_t(r)$ from Eq.~\eqref{eq:background_eqs}, as well as the energy, angular momentum, and orbital frequency of particles in circular orbits using Eqs.~\eqref{eq:enr_ang}.

Our next task is to solve the five coupled first-order ODEs governing the polar sector. These ODEs can be cast in the compact form
\begin{equation}
\frac{d\vec{\psi}}{dr}-\boldsymbol{\alpha}\vec{\psi}=\vec{S}\quad  \ ,
\quad \vec{\psi}=\{K,H_0,H_1,W,\delta \rho\}\ ,\label{eq:masterpolar}
\end{equation}
where $\{K,\,H_0,\, H_1\}$ and $\{W,\,\delta \rho\}$ are the perturbation variables related to the metric and to the fluid, respectively.
The components of the matrix $\boldsymbol{\alpha}$ and of the source vector $\vec{S}$ are given explicitly in Ref.~\cite{Cardoso:2022whc}.  We require the gravitational perturbation variables to satisfy ingoing- and outgoing-wave boundary conditions at the horizon and at our numerical infinity, respectively.  Following Ref.~\cite{Cardoso:2022whc} we assume, for simplicity, that the fluid variables vanish at the two boundaries, i.e. $W(2M_{\text{BH}})=\delta\rho(2M_{\text{BH}})=W(r_\infty)=\delta\rho(r_\infty)=0$.  The gravitational perturbations at the horizon can be expanded as follows:
\begin{equation}
    \begin{aligned}
    \label{eq:polar_hor_BC}
    K^{(\text{in})}(r)&=e^{-i \omega r*}\sum_{i=0}^{N_{\text{in}}}K_H^{(i)}(r-2M_{\text{BH}})^i\ , \\
    H_0^{(\text{in})}(r)&=e^{-i \omega r*}\sum_{i=0}^{N_{\text{in}}}H_{0,H}^{(i)}(r-2M_{\text{BH}})^{i-1}\ , \\
    H_1^{(\text{in})}(r)&=e^{-i \omega r*}\sum_{i=0}^{N_{\text{in}}}H_{1,H}^{(i)}(r-2M_{\text{BH}})^{i-1}\ ,
    \end{aligned}
\end{equation}
where in our numerical calculations we set $N_{\text{in}}=5$.  We similarly expand the background functions at the horizon,
 \begin{equation}
    \begin{aligned}
    m(r)&=M_{\text{BH}}+\sum_{i=2}^{N_{\text{in}}}m_H^{(i)}(r-2M_{\text{BH}})^i\ , \\
    a(r)&=\sum_{i=1}^{N_{\text{in}}}a_H^{(i)}(r-2M_{\text{BH}})^i\ ,
    \end{aligned}
\end{equation}
and we use the numerically computed profiles for $m(r)$ and $a(r)$ to determine the coefficients $a_H^{(i)},\,m_H^{(i)}$.  Given these coefficients, we can expand Eqs.~\eqref{eq:masterpolar} in powers of $r-2M_{\text{BH}}$ and solve order-by-order to find the coefficients $K_H^{(i)}$, $H_{0,H}^{(i)}$, $H_{1,H}^{(i)}$. This completely specifies the boundary conditions at the horizon, modulo the value of $K_H^{(0)}$.

\begin{table*}[t]
\setlength{\tabcolsep}{0.6em}
\begin{tabularx}{\linewidth}{l c | c | c | c | c | c }
\hline \hline $\ell~m$ &\makecell{Vacuum}& &\makecell{Hernquist}&\makecell{NFW\\$r_c=5a_0$}&\makecell{Einasto\\$r_e=a_0$}&\makecell{DM Spike}  \\ \hline \hline 
& & \diag{.2em}{1.1cm}{$\tfrac{M_{\text{Halo}}}{M_{\rm BH}}$}{$\tfrac{M_{\text{Halo}}}{a_0}$} & \makecell{0.1 ~~~~~~~~ 0.001} & \makecell{0.1 ~~~~~~~~ 0.001} & \makecell{0.1 ~~~~~~~~ 0.001} & \makecell{0.1 ~~~~~~~~ 0.001}\\ \hline
$2~2$ & 1.7068e-4 &\makecell{1\\100} &\makecell{1.9727e-4  ~ 1.7029e-4 \\ 1.4045e-4  ~ 1.7035e-4} & \makecell{1.8620e-4  ~ 1.7035e-4 \\ 1.4352e-4  ~ 1.7034e-4} & \makecell{2.9443e-4  ~ 1.6915e-4 \\ 4.7683e-4  ~ 1.6831e-4} & \makecell{1.8033e-4 ~  1.6997e-4 \\ 1.6533e-4  ~ 1.7037e-4}  \\ \hline
$3~1$ & 2.2443e-9 & \makecell{1\\100} & \makecell{5.9746e-9  ~ 2.1793e-9 \\ 2.1349e-9  ~ 2.1742e-9} & \makecell{9.0170e-10 ~ 2.1774e-9 \\ 2.0365e-9  ~ 2.1746e-9} & \makecell{5.8263e-6  ~ 2.3559e-9 \\ 3.0898e-8  ~ 2.1500e-9} & \makecell{5.4913e-9  ~ 2.4933e-9 \\ 3.1469e-9  ~ 2.2568e-9}  \\ \hline
$3~3$ & 2.5488e-5 & \makecell{1\\100} & \makecell{2.8080e-5 ~  2.5421e-5 \\ 2.0893e-5  ~ 2.5441e-5} & \makecell{2.6209e-5  ~ 2.5429e-5 \\ 2.1384e-5 ~  2.5428e-5} & \makecell{2.8465e-5 ~  2.5217e-5 \\ 6.2005e-6  ~ 2.5124e-5} & \makecell{2.5779e-5 ~  2.5323e-5 \\ 2.4470e-5 ~  2.5415e-5}  \\ \hline
$4~2$ & 2.5855e-9 & \makecell{1\\100} &\makecell{1.5745e-8  ~ 2.5000e-9 \\ 1.4164e-9  ~ 2.5099e-9} & \makecell{1.3136e-8  ~ 2.5050e-9 \\ 1.7439e-9 ~  2.5095e-9} & \makecell{8.3112e-10 ~  2.1609e-9 \\ 2.1759e-9  ~ 2.4770e-9} & \makecell{4.6899e-9 ~  2.0384e-9 \\ 8.3514e-10 ~  2.4449e-9}  \\ \hline
$4~4$ & 4.7327e-6 & \makecell{1\\100} &\makecell{4.6286e-6 ~  4.7162e-6 \\ 3.8484e-6  ~ 4.7172e-6} & \makecell{4.3036e-6 ~  4.7176e-6 \\ 3.9524e-6 ~  4.7177e-6} & \makecell{5.7701e-6 ~  4.6656e-6 \\ 9.4731e-7 ~  4.6612e-6} & \makecell{4.3577e-6  ~ 4.6794e-6 \\ 4.4617e-6  ~  4.7165e-6}  \\ \hline \hline
\end{tabularx}
\caption{Selected polar multipolar components $\dot{E}^{\infty}_{(\ell,m)}$ of the energy flux at spatial infinity in units of $q^2=m_p^2/M_{\text{BH}}^2$ for a particle orbiting at $r_p=7.9456M_{\text{BH}}$.
  In each case, we set the width of the Gaussian to $\sigma=1/200$.
  For nonvacuum fluxes, the four entries in each cell span two values of the DM halo mass and compactness: $M_{\text{Halo}}/M_{\text{BH}}=1,\,100$, $M_{\text{Halo}}/a_0=0.1,\,0.001$.
}\label{tab:fluxes}
\end{table*}

We perform a similar procedure at spatial infinity, where we expand the gravitational perturbations as
\begin{equation}
    \begin{aligned}
    \label{eq:polar_inf_BC}
    K^{(\text{out})}(r)&=e^{i \omega r*}\sum_{i=0}^{N_{\text{out}}}\frac{K_\infty^{(i)}}{r^{i}}\ , \\
    H_0^{(\text{out})}(r)&=e^{i \omega r*}\sum_{i=0}^{N_{\text{out}}}\frac{H_{0,\infty}^{(i)}}{r^{i-1}}\ , \\
    H_1^{(\text{out})}(r)&=e^{i \omega r*}\sum_{i=0}^{N_{\text{out}}}\frac{H_{1,\infty}^{(i)}}{r^{i-1}}\ ,
    \end{aligned}
\end{equation}
and the metric functions as
\begin{equation}\label{eq:infma}
    m(r)=\sum_{i=0}^{N_{\text{out}}}\frac{m_\infty^{(i)}}{r^i}\quad\ , \quad
    a(r)=\sum_{i=0}^{N_{\text{out}}}\frac{a_\infty^{(i)}}{r^i}\,.
\end{equation}
Once again we set $N_{\text{out}}=5$.
By matching Eqs.~\eqref{eq:infma} with the numerical profiles of $a(r)$ and $m(r)$ we can determine the values of $a_\infty^{(i)}$ and $m_\infty^{(i)}$.  Then we expand Eqs.~\eqref{eq:masterpolar} in powers of $1/r$ and solve order-by-order for the coefficients $K_\infty^{(i)}$, $H_{0,\infty}^{(i)}$, $H_{1,\infty}^{(i)}$. Fixing $K_\infty^{(0)}=1$ completely specifies the boundary values for the gravitational variables at spatial infinity.

We solve the five ODEs in the frequency domain for $K(r)$, $H_0(r)$, $H_1(r)$, $W(r)$, and $\delta\rho(r)$ using standard \texttt{Mathematica} numerical routines for differential equations. We assume the boundary conditions listed in Eq.~\eqref{eq:polar_hor_BC} at $r=2M_{\text{BH}}$, and those listed in Eq.~\eqref{eq:polar_inf_BC} at $r=r_\infty$.  The source term in Eq.~\eqref{eq:masterpolar} depends on the particle location through a delta function $\delta(r-r_p)$, which we approximate by a Gaussian $\delta(r-r_p) \sim \text{exp}[-(r-r_p)^2/2\sigma^2]/\sqrt{2\pi}\sigma$, as in Refs.~\cite{Figueiredo:2023gas,Cardoso:2022whc}. We also fix the radial and transverse sound speeds which appear in the matrix $\boldsymbol{\alpha}$ of Eq.~\eqref{eq:masterpolar} to be $(c_{s,t},c_{s,r})=(0,9/10)$ (see~\cite{Cardoso:2022whc} for a discussion).
The gravitational perturbations are then extracted at $r_{\text{obs}}=\max[3\times 10^3/\omega,\,2a_0]$.  We make this choice for all the cases we consider, except for profiles with $M_{\text{Halo}}=100$, $a_0=10^5M_{\text{BH}}$, which require a larger extraction radius $r_{\text{obs}}=2\times 10^4/\omega$ to achieve sufficient numerical accuracy.

For a given orbital radius $r_p$, the solution of Eqs.~\eqref{eq:masterpolar} is found by a shooting method, with the shooting parameter $K_H^{(0)}$ determined by requiring that the perturbations and their derivatives must be continuous at some radius $r_\textnormal{obs}$:
\begin{align}
\label{eq:numerical_condition}
\lim_{r\to r_{\text{obs}}} 
\bigg[K_{\text{num}}(r)&K'_{\text{exp}}(r)\nonumber\\
&-K'_{\text{num}}(r)K_{\text{exp}}(r)\bigg]=0\ ,
\end{align} 
In this equation, the quantity $K_{\text{num}}$ refers to the numerically computed $K(r)$, while $K_{\text{exp}}$ refers to the expansion of Eq.~\eqref{eq:polar_inf_BC}.

\begin{figure*}[t]
\label{fig:Flux_comparison_plots_DMspike}
  \centering
  \subfigure{\includegraphics[scale=0.495]{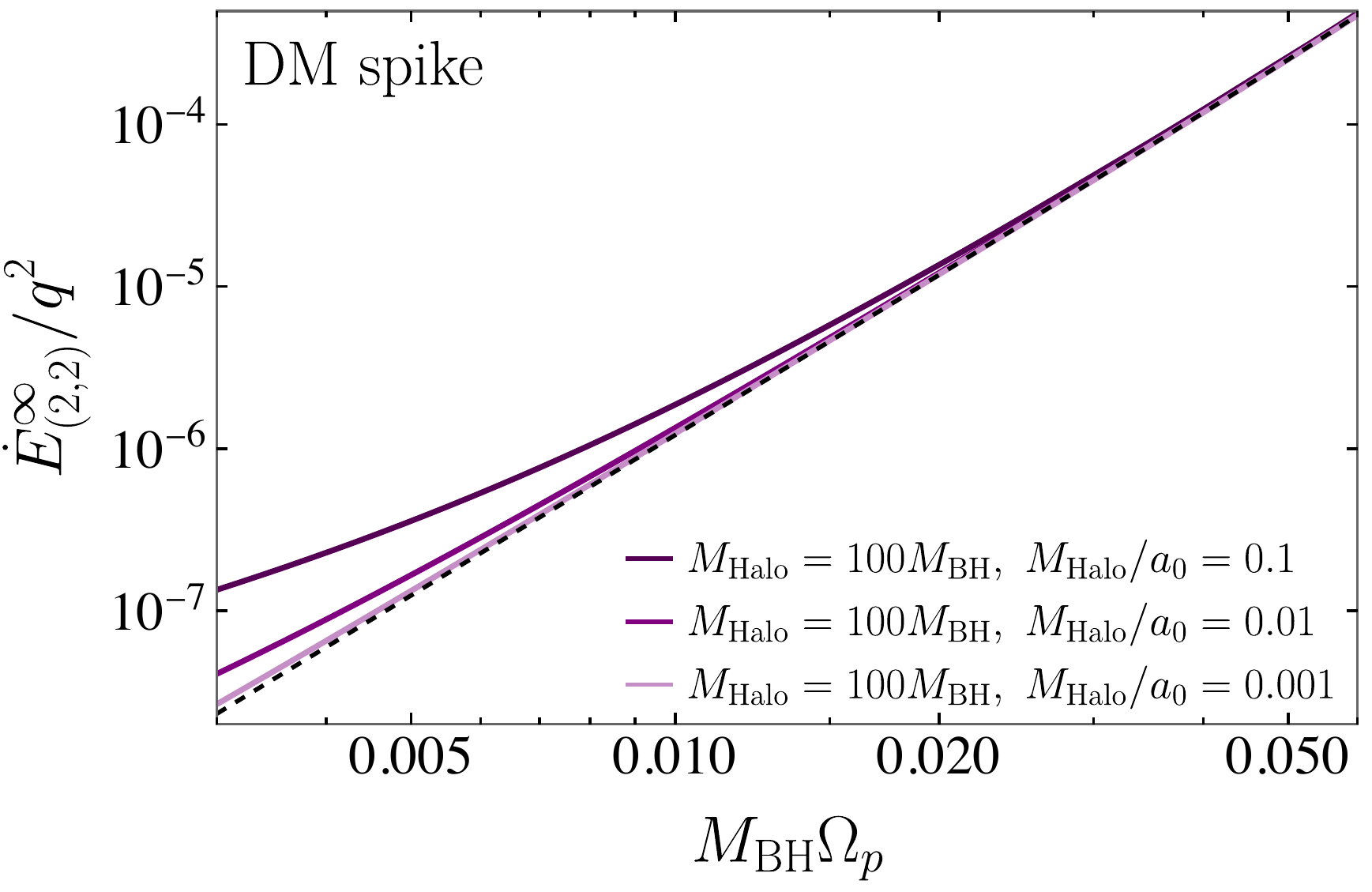}
  \label{fig:DMspike_fluxes_fixed_M}
  }
  \subfigure{\includegraphics[scale=0.495]{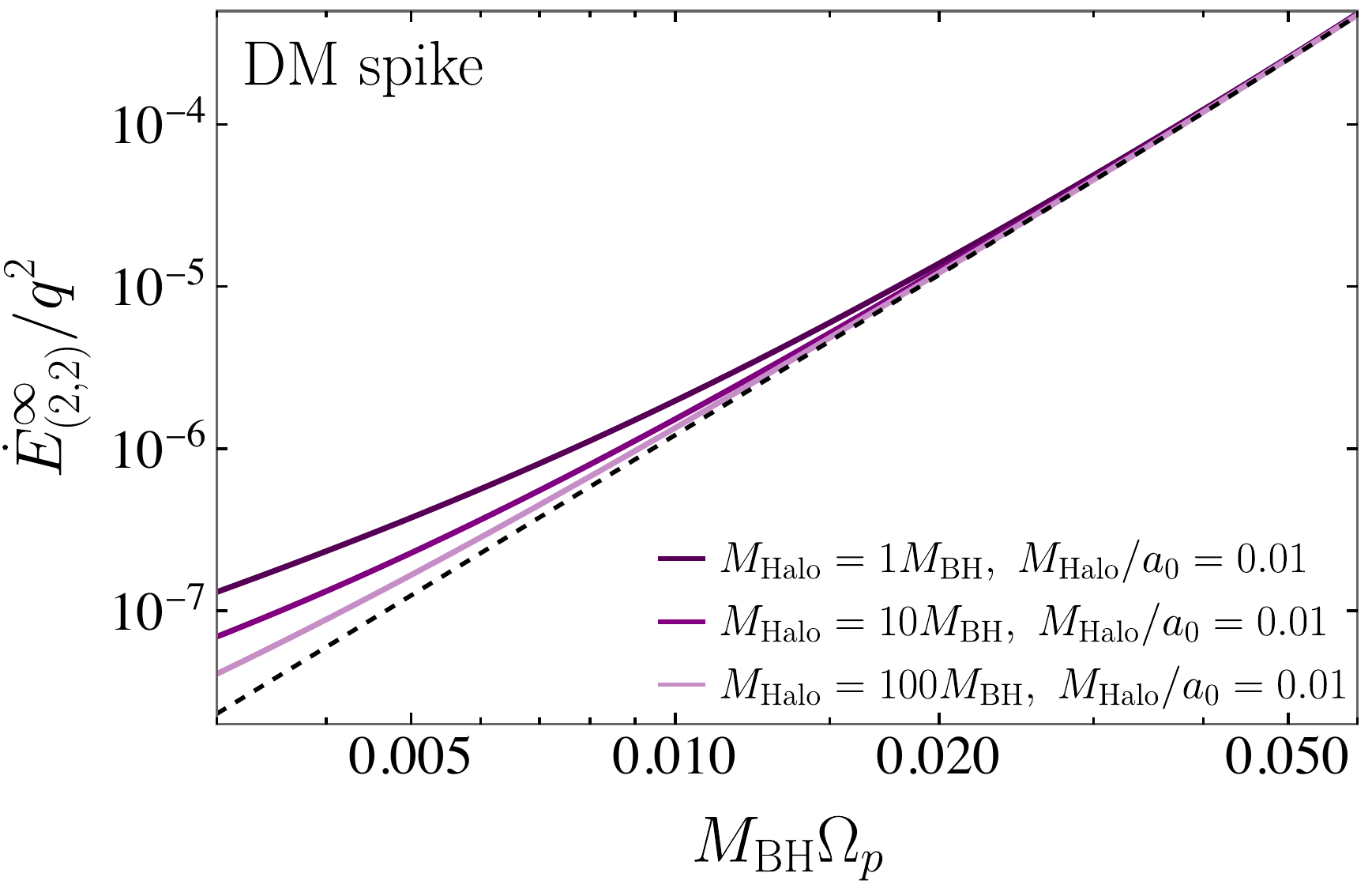} \label{fig:DMspike_fluxes_fixed_compactness}
  }
  \caption{Solid lines: $\ell=m=2$ mode GW energy fluxes, in units of $q^2=m_p^2/M_{\text{BH}}^2$, in the presence of a DM spike matter background. The frequency range on the x axis corresponds to particle orbital radii ranging from $r_p=50M_{\text{BH}}$ to $r_p=6M_{\text{BH}}$. In the left panel we compare DM spike halos with the same value of $M_{\text{Halo}}/M_{\text{BH}}=100$ but different compactness $M_{\text{Halo}}/a_0$, while in the right panel we compare DM spike halos with fixed compactness $M_{\text{Halo}}/a_0=0.01$ but different values of $M_{\text{Halo}}/M_{\text{BH}}$. For reference, the dashed line shows the $\ell=m=2$ multipolar component of the vacuum energy flux.}
\end{figure*}

\begin{figure*}[t]
\label{fig:Flux_comparison_plots_same_compactness}
  \centering
  \subfigure{\includegraphics[scale=0.495]{HQ_Flux_equal_compactness_ref.pdf} \label{fig:Hernquist_fluxes_comparison_same_compactness}
  }
  \subfigure{\includegraphics[scale=0.495]{NFW_Flux_equal_compactness_ref.pdf} \label{fig:NFW_fluxes_comparison_same_compactness}
  }
  \subfigure{\includegraphics[scale=0.495]{Ein_Flux_equal_compactness_ref.pdf} \label{fig:Einasto_fluxes_comparison_same_compactness}
  }
  \subfigure{\includegraphics[scale=0.495]{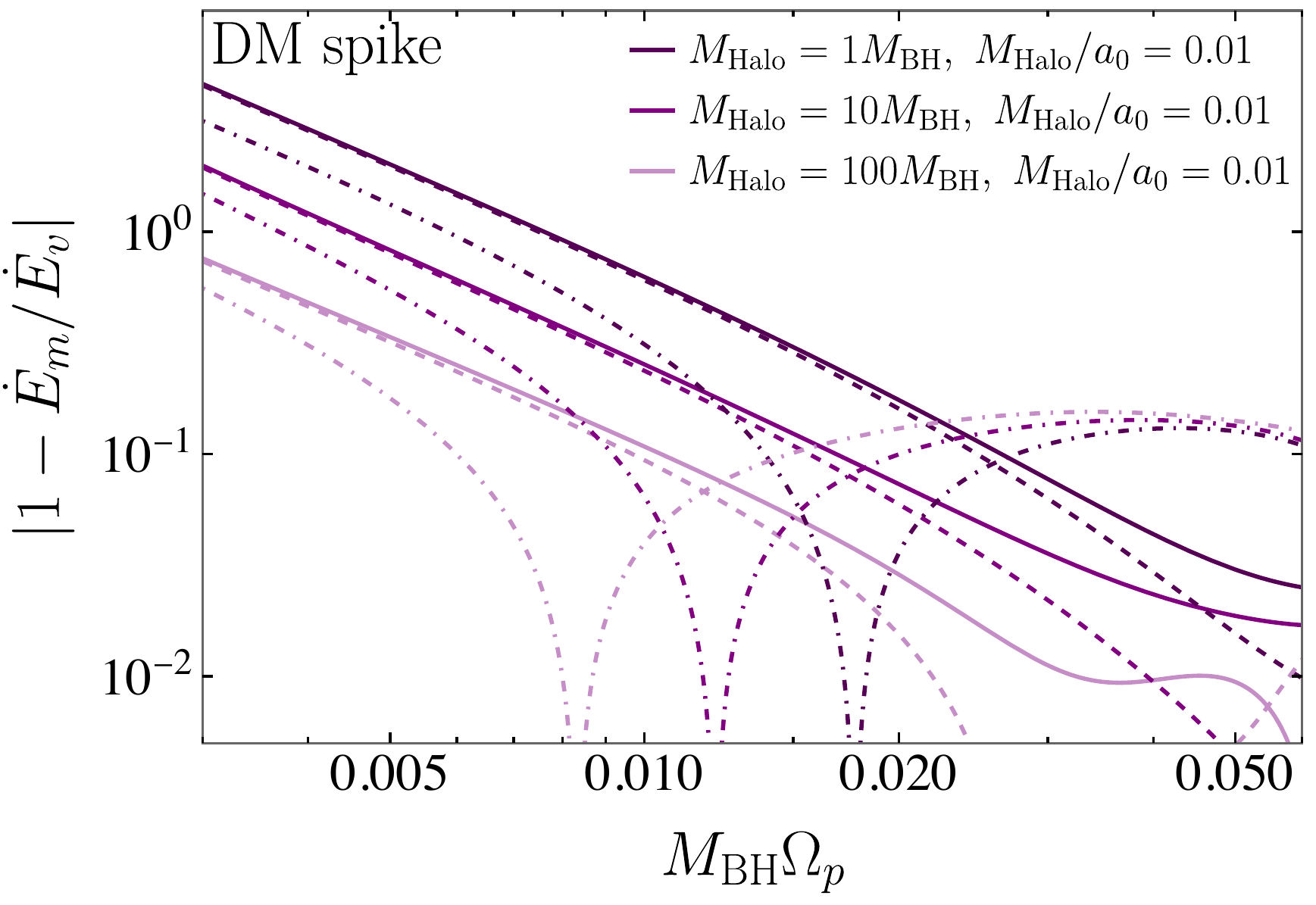} \label{fig:HQspike_fluxes_comparison_same_compactness}
  }
  \caption{Solid lines: relative difference between the $\ell=m=2$ GW energy fluxes, in units of $m_p^2/M_{\text{Halo}}^2$, in vacuum ($\dot{E}_v$) and in the presence of a matter background ($\dot{E}_m$) for different total halo masses, fixing the compactness to $M_{\text{Halo}}/a_0=1/100$. This frequency range corresponds to particle orbital radii ranging from $r_p=50M_{\text{BH}}$ to $r_p=6M_{\text{BH}}$. Dashed lines: relative difference between $\dot{E}_m$ and the vacuum fluxes redshifted according to Eq.~\eqref{eq:redshift}.  Dot-dashed lines: relative difference between $\dot{E}_m$ and the post-Newtonian fluxes of Eq.~\eqref{eq:PN_Edot}.}
\end{figure*}

\begin{figure*}[t]
\label{fig:Flux_comparison_plots_same_Mhalo}
  \centering
  \subfigure{\includegraphics[scale=0.495]{HQ_Flux_equal_Mhalo_ref.pdf} \label{fig:Hernquist_fluxes_comparison_same_Mhalo}
  }
  \subfigure{\includegraphics[scale=0.495]{NFW_Flux_equal_Mhalo_ref.pdf} \label{fig:NFW_fluxes_comparison_same_Mhalo}
  }
  \subfigure{\includegraphics[scale=0.495]{Ein_Flux_equal_Mhalo_ref.pdf} \label{fig:Einasto_fluxes_comparison_same_Mhalo}
  }
  \subfigure{\includegraphics[scale=0.495]{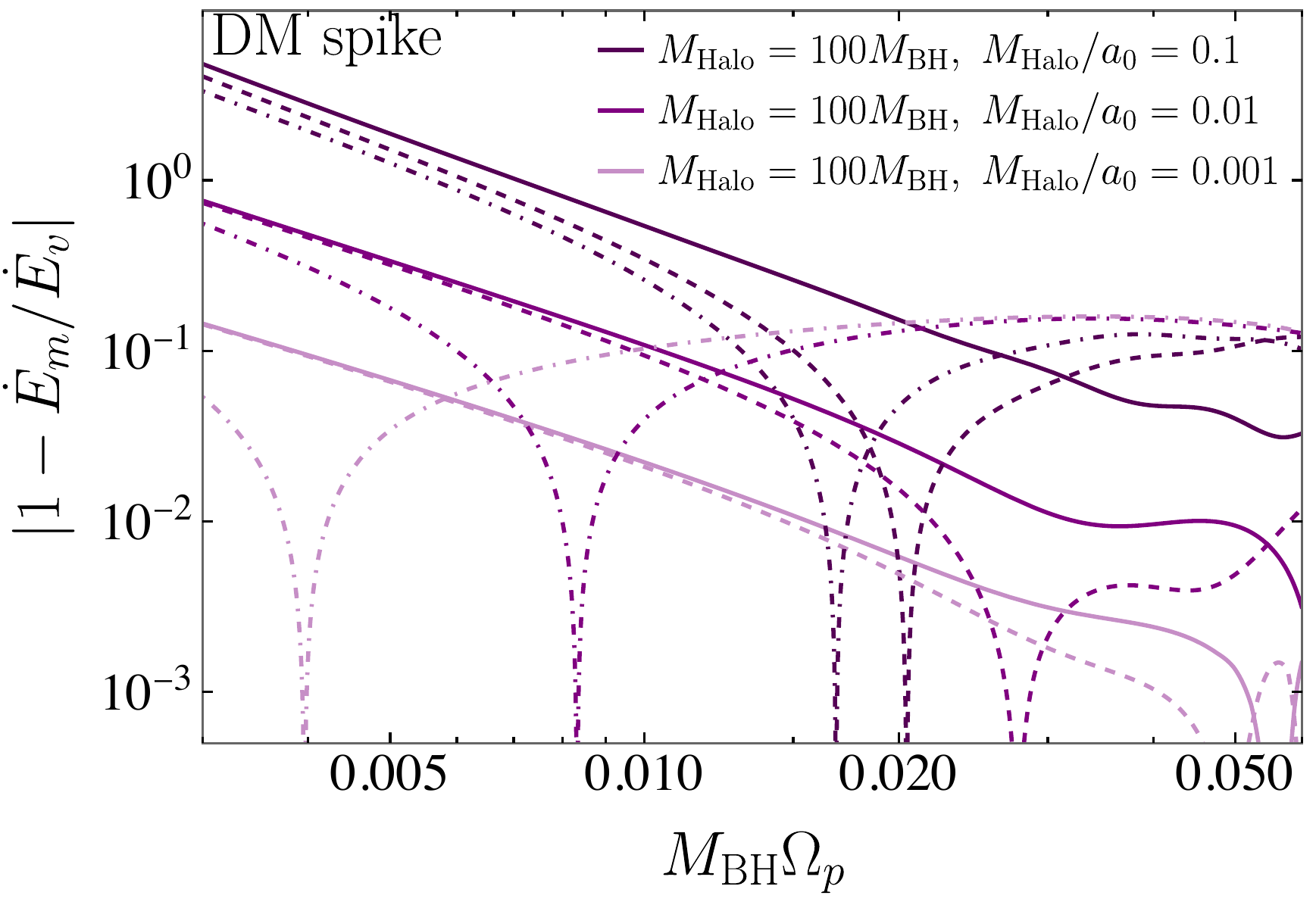} \label{fig:HQspike_fluxes_comparison_same_Mhalo}
  }
  \caption{Solid lines: relative difference between the $\ell=m=2$ GW energy fluxes in vacuum ($\dot{E}_v$) and in the presence of matter ($\dot{E}_m$) for different compactness, fixing the halo mass to $M_{\text{Halo}}=100M_{\text{BH}}$. This frequency range corresponds to particle orbital radii ranging from $r_p=50M_{\text{BH}}$ to $r_p=6M_{\text{BH}}$. Dashed lines: relative difference between $\dot{E}_m$ and the vacuum fluxes redshifted according to Eq.~\eqref{eq:redshift}.  Dot-dashed lines: relative difference between $\dot{E}_m$ and the post-Newtonian fluxes of Eq.~\eqref{eq:PN_Edot}.}
\end{figure*}

We can finally compute the GW fluxes at infinity for different multipolar components $(\ell,\,m)$ from the relation
\begin{equation} \dot{E}^{\infty}_{(\ell,m)}/q^2=\lim_{r\to r_{\text{obs}}}\frac{1}{32\pi}\frac{(\ell+2)!}{(\ell-2)!}|K(r)|^2\,,~~ (\ell+m ~\text{even})\,,
\end{equation}
where $q=m_p/M_{\text{BH}}$.
The codes used to calculate the background quantities and the polar fluxes are publicly available online~\cite{SGREP_REPO}.

\section{Results}
\label{sec:Results}

Examples of the GW fluxes $\dot{E}^{\infty}_{(\ell,m)}$ computed in this way are shown in Table~\ref{tab:fluxes}.
All fluxes are evaluated at a reference particle orbital radius $r_p=7.9456M_{\text{BH}}$ to facilitate comparison with the vacuum fluxes computed in Ref.~\cite{Martel:2003jj}.
For nonvacuum fluxes and for a given DM profile, the four entries in each cell of the table span two values of the DM halo mass and compactness: $M_{\text{Halo}}/M_{\text{BH}}=1,\,100$ and $M_{\text{Halo}}/a_0=0.1,\,0.001$. In other words, we consider the four most extreme cases shown in Fig.~\ref{fig:Density_profiles_grid}: the top (bottom) entries in each cell correspond to a small (large) halo mass, while the left (right) entries correspond to a large (small) compactness.

Let us first focus on halos with small compactness, $M_{\text{Halo}}/a_0=0.001$ (right entries in each cell of Table~\ref{tab:fluxes}). In this case, changes in the halo mass $M_{\text{Halo}}/M_{\text{BH}}$ have nearly no effect: in fact the fluxes are very similar for any choice of the DM halo profile, and they differ by at most a few percent from the corresponding vacuum fluxes. This makes sense when we look at the DM distributions plotted in the right panels of Fig.~\ref{fig:Density_profiles_grid}: if the compactness is small, most of the DM density is at large radii. The particle evolution is most sensitive to the DM density inside the orbital radius (marked by vertical lines in Fig.~\ref{fig:Density_profiles_grid}), which is always very low for ``dilute'' DM profiles with $M_{\text{Halo}}/a_0=0.001$, with peak values of $M_{\text{BH}}^2\rho\lesssim 10^{-6}$. Therefore, the DM distribution close to the BH has an almost negligible effect on the GW energy flux and on the orbital evolution of the particle.

Consider next halos with large compactness, $M_{\text{Halo}}/a_0=0.1$ (left entries in each cell of Table~\ref{tab:fluxes}). We now observe that the multipolar contributions to the GW energy fluxes are typically larger for small values of the halo mass ($M_{\text{Halo}}/M_{\text{BH}}=1$) than for large values of the halo mass ($M_{\text{Halo}}/M_{\text{BH}}=100$). Once again, this can be understood by looking at the DM distributions plotted in the left panels of Fig.~\ref{fig:Density_profiles_grid}. Consider for reference the peak of the Einasto DM density distribution (dotted green line): it decreases from $M_{\text{BH}}^2\rho\sim 10^{-3}$ for $M_{\text{Halo}}/M_{\text{BH}}=1$, to
$M_{\text{BH}}^2\rho\sim 10^{-4}$ for $M_{\text{Halo}}/M_{\text{BH}}=100$. This is because {\em at fixed compactness}, the more massive halos are more diluted, and extend out to larger orbital radii, so the amount of DM {\it within the particle orbit} decreases as the total halo mass $M_{\text{Halo}}/M_{\text{BH}}$ increases. This explains why, at fixed compactness, the multipolar contributions to the GW energy fluxes are typically smaller for larger values of the halo mass.
There are some exceptions to this general rule, such as the $(\ell,\,m)=(3,\,1)$ multipole for the NFW profile and the $(\ell,\,m)= (4,\,4)$ multipole for the DM spike, but as we will discuss below these are largely subdominant and, ultimately, they do not contribute much to the energy fluxes and to the orbital evolution of the particle.

In Fig.~\ref{fig:Flux_comparison_plots_DMspike} we focus on the fiducial ``DM spike'' model, which was computed from a fully relativistic calculation of the adiabatic growth of a Hernquist profile~\cite{Sadeghian:2013laa,Speeney:2022ryg}. The trends visible in this figure are consistent with Table~\ref{tab:fluxes}.
The left panel shows that the quadrupolar energy flux decreases as the DM halo become less compact at fixed values of $M_{\text{Halo}}/M_{\text{BH}}$: this is easy to understand, because the strong-field dynamics of the particle is dominated by the BH when the DM profile is more dilute (i.e., when the DM density close to the BH decreases). The right panel shows that the quadrupolar energy flux is smaller for larger values of $M_{\text{Halo}}/M_{\text{BH}}$ at fixed compactness, again because more dilute DM profiles have less DM within the particle orbit. These are the same trends we observed at $r_p=7.9456M_{\text{BH}}$ in Table~\ref{tab:fluxes}, and they apply to all values of $r_p/M_{\text{BH}}$ (or $M_{\text{BH}}\Omega_p$). 

However, Fig.~\ref{fig:Flux_comparison_plots_DMspike} allows us to better appreciate the dependence of these effects on the particle radius $r_p/M_{\text{BH}}$.
For example, the right panel shows that the effect of gravitational redshift is still important. When we fix $M_{\text{Halo}}/a_0=0.01$, $a_0$ changes by two orders of magnitude as we vary $M_{\text{Halo}}/M_{\text{BH}}$, so the emitted GWs propagate through very different density profiles.
For $r_p=50 M_{\text{BH}}$ ($M_{\text{BH}} \Omega_p\sim 0.003$) the difference between the three cases is more pronounced, because the particle is (roughly) located halfway through the DM distribution when $M_{\text{Halo}}/M_{\text{BH}}=1$ (so that $a_0=100 M_{\text{Halo}}=100 M_{\text{BH}}$). On the contrary, when $r_p=6 M_{\text{BH}}$ ($M_{\text{BH}} \Omega_p\sim 0.05$), the particle is located so deep within the DM distribution that any differences induced by gravitational redshift saturate, and the three fluxes become nearly indistinguishable.

The solid lines in the top-left panel of Fig.~\ref{fig:Flux_comparison_plots_same_compactness} show the relative difference between the $\ell=m=2$ multipolar contribution to the energy flux in the frequency domain for the Hernquist distribution and the corresponding vacuum result for different DM halos with fixed compactness $M_\textnormal{Halo}/a_0=0.01$. The plot spans values of the orbital frequencies such that the circular orbit radius of the particle varies between $r_p=50M_{\text{BH}}$ and $r_p=6M_{\text{BH}}$. We have checked that when we use a cutoff factor of $(1-2M_{\text{BH}}/r)$ our results are in excellent agreement with those in Fig.~3 of Ref.~\cite{Cardoso:2022whc}, which made use of closed-form analytic expressions for the metric functions $m(r)$ and $a(r)$. In fact, the energy fluxes shown here are smoother compared to those of Ref.~\cite{Cardoso:2022whc}, mainly because of the larger value of $r_{\text{out}}$ considered in this paper to increase numerical accuracy.

In the remaining panels of Fig.~\ref{fig:Flux_comparison_plots_same_compactness} we compare vacuum and matter fluxes for the NFW, Einasto, and DM spike profiles.
The solid lines in different panels shows trends consistent with the DM profiles plotted in Fig.~\ref{fig:Density_profiles_grid}. For example, by comparing the solid lines in the two top panels we observe minimal differences between the Hernquist and NFW fluxes, which is hardly surprising given the similarity of the corresponding DM profiles. The GW flux in the presence of DM differs more from the vacuum case when we consider the DM spike and (even more so) the Einasto profiles. The reason can be easily understood by a glance at the middle column of Fig.~\ref{fig:Density_profiles_grid}: the DM density in the regime of interest is highest for the Einasto profile, followed by the DM spike profile and (at some distance) by the Hernquist and NFW profiles.

The dashed lines in each panel of Fig.~\ref{fig:Flux_comparison_plots_same_compactness} show the relative difference between the numerical values of $\dot{E}^\infty_{(2,2)}$ and the redshifted vacuum fluxes obtained through the scaling~\cite{Cardoso:2022whc}
\begin{equation}
\label{eq:redshift}
\Omega_p\rightarrow \Omega_p/\gamma\quad\ ,\quad 
\omega \rightarrow \omega/\gamma\quad \ ,\quad 
m_p\rightarrow \gamma m_p\ ,
\end{equation}
where the redshift factor is given by $\gamma=1-M_{\text{Halo}}/a_0$ for Hernquist, NFW, and DM spike distributions, and by $\gamma=1-M_{\text{Halo}}/r_e$ for the Einasto model.
Interestingly, the nonvacuum fluxes are not well captured by the scaling~\eqref{eq:redshift}, even for small
values of the halo mass $M_{\text{Halo}}$.
This is a key difference with respect to the energy fluxes in the axial sector: in that case, this simple redshift scaling turned out to significantly reduce the difference between the vacuum energy fluxes and those in the presence of matter~\cite{Cardoso:2022whc,Figueiredo:2023gas}.

Finally, the dot-dashed lines in each panel of Fig.~\ref{fig:Flux_comparison_plots_same_compactness} show the relative difference between the numerical fluxes and the corrections to the vacuum PN GW fluxes computed in Ref.~\cite{Speeney:2022ryg}, i.e.,
\begin{equation}
\label{eq:PN_Edot}
\dot{E}^\infty_{\text{GW}}=\frac{32}{5}(M_{\text{BH}}\Omega_p)^{10/3}\left(1+\frac{4}{3} \epsilon q(r)\right)\ .
\end{equation}
Equation~\eqref{eq:PN_Edot} is obtained
by expanding the energy balance equation in powers of the small quantity $\epsilon q(r)=M_{\text{DM}}(r)/M_{\text{BH}}$, where $M_{\text{DM}}(r)$ is the mass of the halo enclosed in a sphere of radius $r$. Therefore, not surprisingly, the approximation works best when $M_{\text{Halo}}/M_{\text{BH}}$ is smallest. Note that at variance with Ref.~\cite{Speeney:2022ryg}, in Eq.~\eqref{eq:PN_Edot} we do not include dynamical friction terms,
because our framework does not accommodate these effects at leading order in the perturbations. In fact, the agreement with our numerical fluxes gets worse when we compare them against PN expressions which include dynamical friction.

Figure~\ref{fig:Flux_comparison_plots_same_Mhalo} is similar to Fig.~\ref{fig:Flux_comparison_plots_same_compactness}, except now we fix the ratio $M_{\text{Halo}}/M_{\text{BH}}=100$ and vary the halo compactness: $M_{\text{Halo}}/a_0=0.1, 0.01, 0.001$. This plot confirms the general trends observed earlier. The Hernquist and NFW energy fluxes behave in very similar ways. The Einasto and (to a lesser extent) the DM spike profile produce much larger deviations from the vacuum fluxes. Finally, and most importantly, the dephasing induced by DM generally increases with the compactness of the halo.

Figures~\ref{fig:Flux_comparison_plots_same_compactness} and \ref{fig:Flux_comparison_plots_same_Mhalo} show that Eq.~\eqref{eq:PN_Edot} is not sufficient to reproduce the numerical fluxes at the orbital frequencies of interest for EMRIs for any of the DM profiles we consider. The agreement between Eq.~\eqref{eq:PN_Edot} and $\dot{E}^\infty_{2,2}$ improves at large radii and small orbital frequencies, as expected in any PN treatment, but it gets worse for more astrophysically realistic (i.e., less compact) halos.

Overall, these results lead to two main conclusions: (i) the polar fluxes cannot be adequately described by PN expansions, and they cannot be simply attributed to gravitational redshift effects; (ii) the degeneracies due to the redshift scaling \eqref{eq:redshift} found in the axial sector are generally broken when we consider polar perturbations. This observation paves the way for the interesting possibility of constraining the halo properties through GW observations with space-based detectors.

\begin{figure}[t]
  \centering
  \includegraphics[width=\linewidth]{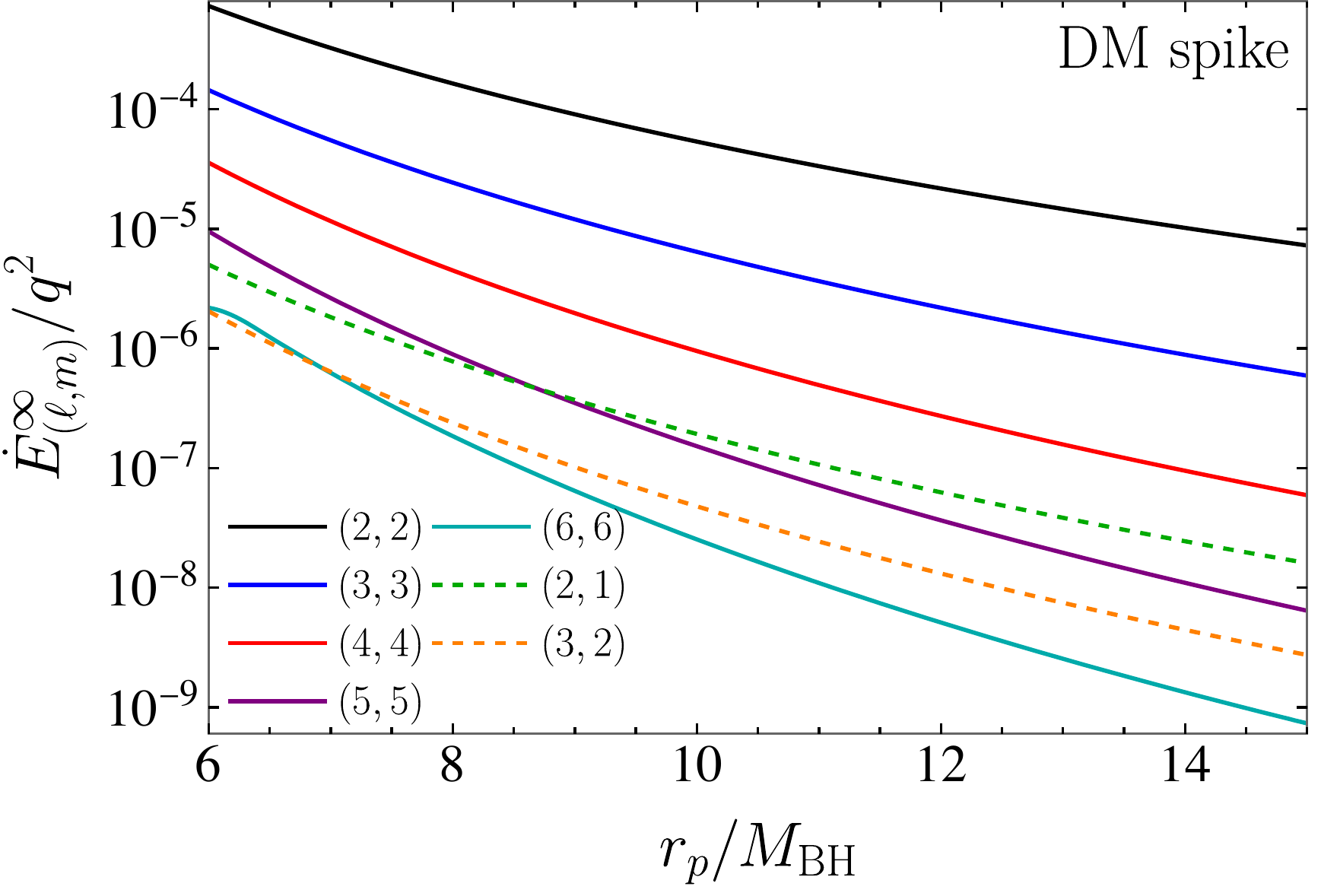}
  \caption{Multipolar components of the GW energy flux in the presence of a DM spike background as a function of the particle's orbital radius $r_p/M_{\text{BH}}$. We fix the halo mass to $M_{\text{Halo}}/M_{\text{BH}}=100$ and the compactness to $M_{\text{Halo}}/a_0=0.001$. Solid (dashed) lines refer to polar (axial) modes.}
\label{fig:DMspike_multipoles}
\end{figure}

Finally, in Fig.~\ref{fig:DMspike_multipoles} we show the multipolar components of the GW energy flux as a function of the orbital radius $r_p/M_{\text{BH}}$ for a DM spike model with $M_{\rm Halo}/a_0=0.001$ and $M_{\rm Halo}=100M_{\rm BH}$. For completeness we also include the $(2,\,1)$ and $(3,\,2)$ axial contributions, that were first computed in Ref.~\cite{Figueiredo:2023gas}. Similarly to the vacuum case, we observe a clear hierarchy in the amplitude of the GW multipoles. The total energy flux is dominated by the $(2,\,2)$, $(3,\,3)$ and $(4,\,4)$ modes. Among the subdominant components, the axial $(2,\,1)$ multipole dominates over the polar $(5,\,5)$ multipole at large values of $r_p/M_{\text{BH}}$, while the opposite is true for $r_p/M_{\text{BH}}\lesssim 8.5$. We observe an analogous behavior for the axial $(3,\,2)$ and polar $(6,\,6)$ modes, with a ``crossing radius'' around $r_p/M_{\rm BH}\sim 7$.  Similar considerations apply to the Hernquist, NFW, and Einasto models, which are not shown here for brevity.

\section{Conclusions}
\label{sec:conclusion}
This work extends the fully relativistic modeling of EMRIs embedded in a DM environment to include a numerical treatment of the polar sector, thus completing the program initiated in Refs.~\cite{Cardoso:2022whc,Figueiredo:2023gas}.

We have used this formalism to study the GW fluxes produced by different DM density distribution models, and we have found that the differences can be of order unity or larger with respect to the Schwarzschild fluxes, depending on our assumptions on the DM density profile.

Some general trends can be identified from Table~\ref{tab:fluxes}.
For halos with large compactness (say, $M_{\text{Halo}}/a_0=0.1$), the multipolar contributions to the GW energy fluxes are typically large for {\em small} values of the halo mass ($M_{\text{Halo}}/M_{\text{BH}}=1$), because at fixed compactness the more massive halos are more diluted, extending out to larger orbital radii, so the amount of DM within the particle orbit decreases as $M_{\text{Halo}}/M_{\text{BH}}$ increases.
In the more realistic case of halos with small compactness, $M_{\text{Halo}}/a_0\ll 1$, the fluxes are very similar for any choice of the DM halo profile and for any value of the halo mass $M_{\text{Halo}}/M_{\text{BH}}$, differing by at most a few percent from the corresponding vacuum fluxes. This is because most of the DM density is at large radii, and therefore the DM distribution close to the BH has an almost negligible effect on the GW energy flux and on the orbital evolution of the particle.

In Fig.~\ref{fig:Flux_comparison_plots_DMspike} we focus on the fiducial ``DM spike'' model, computed from a fully relativistic calculation of the adiabatic growth of a BH within the Hernquist profile~\cite{Sadeghian:2013laa,Speeney:2022ryg}.
The quadrupolar energy flux decreases as the DM halo become less compact at fixed values of $M_{\text{Halo}}/M_{\text{BH}}$, because the strong-field dynamics of the particle is dominated by the BH when the DM profile is more dilute (i.e., when the DM density close to the BH decreases). The energy flux is smaller for large values of $M_{\text{Halo}}/M_{\text{BH}}$ at fixed compactness for a similar reason: more dilute DM profiles have less DM within the particle orbit.
Realistic DM profiles are probably in the regime $M_{\text{Halo}}/a_0\ll 1$, as is the case for a Milky Way-like galaxy. We plan to explore more carefully this ``dilute DM'' limit in future work.

As shown in Figs.~\ref{fig:Flux_comparison_plots_same_compactness} and \ref{fig:Flux_comparison_plots_same_Mhalo}, the numerical energy fluxes in the polar case cannot be captured by simply taking into account the gravitational redshift of the vacuum fluxes (at variance with the axial case), nor by leading-order corrections to the mass profile combined with PN estimates of the flux.

Our analysis has several limitations that we plan to address in future studies.

First of all, it is important to accurately model dynamical friction, which may accumulate large dephasings over the $\mathcal{O}(10^5-10^6)$ orbital cycles typical of LISA EMRIs~\cite{Eda:2014kra,Cardoso:2019rou,Speeney:2022ryg,Coogan:2021uqv,Cardoso:2019mqo,Traykova:2021dua,Traykova:2023qyv}.
Moreover, our model does not properly take into account matter fluxes. While EMRIs can be correctly described by stationary configurations with matter profiles that vanish at the horizon and at spatial infinity, intermediate-mass binaries will be sensitive to feedback effects. This backreaction will affect the DM profile during the evolution of the orbit~\cite{Kavanagh:2020cfn,Coogan:2021uqv,Nichols:2023ufs}. A proper treatment of backreaction requires the inclusion of higher-order terms in the perturbative expansion. If  simple estimates similar to Eq.~\eqref{eq:PN_Edot} hold at very small orbital frequencies, DM halos could produce effects of phenomenological relevance for pulsar timing arrays~\cite{Ghoshal:2023fhh,Aghaie:2023lan}.

On the data analysis side, prior works have proposed to use EMRI observations to distinguish various environmental signatures~\cite{Cole:2022fir,Cole:2022ucw,Barausse:2014tra,Barausse:2014pra,Speeney:2022ryg, Vicente:2022ivh,Rahman:2023sof}. The majority of these works use a combination of fully relativistic and PN estimates of environmental effects, but so far there is no consistent description of the evolution of the BH spacetime embedded in DM during the inspiral.  Our framework can accommodate relativistic effects in the generation and propagation of GWs. In the future, we plan to develop waveform models based on these flux calculations and to use them in a Bayesian framework to estimate the detectability of DM overdensities.

\acknowledgments
N.S. and E.B. are supported by NSF Grants No. AST-2006538, PHY-2207502, PHY-090003 and PHY-20043, by NASA Grants No.~20-LPS20-0011 and 21-ATP21-0010, by the John Templeton Foundation Grant 62840, and by the Simons Foundation.
A.M. and E.B. acknowledge support from the Indo-US Science and Technology Forum through the Indo-US Centre for Gravitational-Physics and Astronomy, grant IUSSTF/JC-142/2019. This work was supported in part by the Italian Ministry of Foreign Affairs and International Cooperation”, grant number PGR01167. 
We acknowledge support by VILLUM Foundation (grant no. VIL37766) and the DNRF Chair program (grant no. DNRF162) by the Danish National Research Foundation.
V.C.\ is a Villum Investigator and a DNRF Chair.  
V.C. acknowledges financial support provided under the European Union’s H2020 ERC Advanced Grant “Black holes: gravitational engines of discovery” grant agreement no. Gravitas–101052587. 
Views and opinions expressed are however those of the author only and do not necessarily reflect those of the European Union or the European Research Council. Neither the European Union nor the granting authority can be held responsible for them.
This project has received funding from the European Union's Horizon 2020 research and innovation programme under the Marie Sklodowska-Curie grant agreement No 101007855 and No 101131233.
This work was carried out at the Advanced Research Computing at Hopkins (ARCH) core facility (\url{rockfish.jhu.edu}), which is supported by the NSF Grant No.~OAC-1920103.

\bibliography{refs}

\end{document}